\documentclass{aa}  
\usepackage[utf8]{inputenc}
\usepackage{graphicx}
\usepackage{txfonts}
\usepackage{cleveref}
\usepackage{wasysym}
\usepackage{xcolor}
\usepackage{hyperref}   
                                                             
\usepackage{xcolor}                                                             

\crefname{section}{§}{§§}

\begin{document} 

\title{Modelling of proton acceleration in application to a ground level enhancement}

\author{A. Afanasiev\inst{\ref{inst1}} \and R. Vainio\inst{\ref{inst1}} \and A. P. Rouillard\inst{\ref{inst2}, \ref{inst3}} \and M. Battarbee\inst{\ref{inst4}, \ref{inst5}} \and A. Aran\inst{\ref{inst6}} \and P. Zucca\inst{\ref{inst7}}
          }
          
\institute{Department of Physics and Astronomy, University of Turku, Turku, Finland \email{alexandr.afanasiev@utu.fi}\label{inst1}
\and 
Institut de Recherche en Astrophysique et Plan\'{e}tologie, Universit\'{e} de Toulouse III (UPS), France\label{inst2}            
\and
Centre National de la Recherche Scientifique, UMR 5277, Toulouse, France\label{inst3}
\and
Department of Physics, University of Helsinki, Helsinki, Finland\label{inst4} 
\and
Jeremiah Horrocks Institute, University of Central Lancashire, Preston, United Kingdom\label{inst5} 
\and
Departament de Física Quàntica i Astrofísica, Institut de Ciències del Cosmos (ICCUB), Universitat de Barcelona, Barcelona, Spain\label{inst6}                
\and
ASTRON, Netherlands Institute for Radio Astronomy, Postbus 2, 7990 AA, Dwingeloo, The Netherlands\label{inst7}   
}

\date{}
   
\titlerunning{Modelling of proton acceleration in a ground level enhancement}
\authorrunning{Afanasiev et al.}   

 
  \abstract
   {The source of high-energy protons (above \textasciitilde 500~MeV) responsible for the so-called ground level enhancements (GLEs) remains an open question in solar physics. One of the candidates is a shock wave driven by a coronal mass ejection, which is thought to accelerate particles via diffusive-shock acceleration.}
   {We perform physics-based simulations of proton acceleration using information on the shock and ambient plasma parameters derived from the observation of a real GLE event. We analyse the simulation results with the aim to find out which of the parameters are significant in controlling the acceleration efficiency and to get a better understanding of the conditions under which the shock can produce relativistic protons.}
   {We use results of the recently developed technique to determine the shock and ambient plasma parameters, applied to the 17 May 2012 GLE event, and carry out proton acceleration simulations with the Coronal Shock Acceleration model.}
   {We have performed proton acceleration simulations for nine individual magnetic field lines characterised by various plasma conditions. Analysis of the simulation results shows that the acceleration efficiency of the shock, i.e., its ability to accelerate particles to high energies, tends to be higher for those shock portions that are characterised by larger values of the scattering-centre compression ratio $r_\mathrm{c}$ and/or the fast-mode Mach number $M_\mathrm{FM}$. At the same time, the acceleration efficiency can be strengthened due to enhanced plasma density in the corresponding flux tube. The simulations show that protons can be accelerated to GLE energies in the shock portions characterised by the highest values of $r_\mathrm{c}$. Analysis of the delays between the flare onset and the production times of protons of 1~GV rigidity for different field lines in our simulations, and a subsequent comparison of those with the observed values indicate a possibility that quasi-perpendicular portions of the shock play the main role in producing relativistic protons.}
   {}


    \maketitle
%

\section{Introduction}

It is established that enhancements in high-energy particle fluxes from the Sun, known as solar energetic particle (SEP) events, are associated with solar flares and coronal mass ejections (CMEs). Energetic particles associated with CMEs are thought to be accelerated in CME-driven shock waves by means of diffusive shock acceleration (DSA) mechanism \citep[e.g.,][]{Reames-2013}. CME-driven shocks are also considered to be a possible source of relativistic protons responsible for the so-called ground level enhancements (GLEs) \citep[e.g.,][]{Reames09, SandroosVainio09, Gopal12}. For this to be possible, such a shock has to exist and accelerate protons efficiently before the release time of relativistic protons from the Sun. 

One of the observable signatures of a CME-driven shock in the corona is radio emission of type~II detected at meter wavelengths, which is produced by shock-accelerated electrons. \citet{Reames09} in his study of GLE events in solar cycle 23 compared the solar particle release (SPR) times resulting from the velocity dispersion analysis of SEPs with the start times of the corresponding metric type~II radio bursts. He found that all SPR times occurred after the start of metric type~II bursts. \citet{Gopal12} obtained the same result using neutron monitor data to infer the relativistic proton release times. They  concluded that in all of the events shocks existed already before the relativistic particle release. However, observing type~II bursts is not enough to conclude that the shock is able to accelerate protons. To do that, detailed information on the shock parameters is needed.          

Recently, \citet{Rouillard-2016} have computed the distribution of the (fast-mode) Mach number over the CME-driven pressure front in the 17 May 2012 GLE event. They found that a region of increased Mach number was developing with time over the expanding front. This region evolved to a supercritical shock by the SPR time. Having become supercritical, a shock is able to reflect upstream ions that can be then injected to the acceleration process. This, in turn, can lead to an increase in the acceleration efficiency of the shock. For instance, in the time-dependent DSA model involving self-generated turbulence, the maximum particle energy achieved in a given time increases with the particle injection efficiency of the shock \citep[e.g.,][]{Vainio-2014, Afanasiev-2015}.

Thus, the result of \citet{Rouillard-2016} is encouraging for the shock hypothesis of the GLE origin. However, accurate modelling of particle acceleration is needed in order to give a definitive answer. Present efforts are aimed at coupling models of particle acceleration (and transport) with realistic semi-empirical/magnetohydrodynamic (MHD) models of the CME lift-off  \citep[e.g.,][]{Kozarev-2013, Kozarev-2016}. This is important for better understanding under what conditions high-energy particle production occurs and which ambient plasma and shock parameters mainly control the acceleration efficiency of the shock. In particular, \citet{Kozarev-2016} combined an analytic test-particle model of DSA with semi-empirical modelling of the evolving shock front. Their modelling shows that the most efficient particle acceleration takes place in highly oblique (quasi-perpendicular) portions of the shock. Other important parameters are the shock speed and gas compression ratio.   

In this paper, we take the same approach to combine a model of DSA with a semi-empirical shock model. However, our DSA model is not a test-particle model but accounts self-consistently for the formation of Alfv\'enic turbulent foreshock. Correspondingly, it addresses parallel and oblique shocks. Regarding the semi-empirical shock model, we use the one obtained by \citet{Rouillard-2016} for the CME-driven shock associated with the SEP/GLE event of 17 May 2012. The goals of this work are 1) to identify which shock and plasma parameters are mainly responsible for efficient particle acceleration in oblique shocks in realistic conditions, and 2) to try to get a better understanding of the production conditions at the shock of relativistic protons in this GLE event.    

Note that the generation of Alfv\'enic turbulence in coronal shocks was addressed previously in a number of papers, e.g., \citet{Vainio-2007, Vainio-2008}, \citet{NgReames08}, \citet{Afanasiev-2015} (see also \citealt{NgReames94} and \citealt{Vainio03}). In particular, \citet{NgReames08} demonstrated that a fast parallel shock under typical coronal conditions is able to accelerate low-energy seed protons into a power-law energy spectrum with the cutoff energy of 300 MeV in \textasciitilde10 minutes. However, as it follows from our study, in order to resolve current uncertainties concerning GLEs, it is important to conduct particle acceleration modelling for the event-specific conditions and to compare the results  with the event-specific observations.        

The paper is organised as follows. In \cref{sec:methods}, the numerical model to simulate particle acceleration as well as the techniques used to derive the ambient plasma and shock parameters are briefly described. In \cref{sec:results} and \cref{sec:discussion}, simulation results and their discussion are presented. In \cref{sec:conclusions}, conclusions are provided.        	 

\section{Methods}\label{sec:methods}
\subsection{Particle acceleration simulation}
   
In this work, we employ the Coronal Shock Acceleration (CSA) code for particle acceleration simulations. It is a Monte Carlo simulation model of diffusive-shock acceleration of ions in coronal and interplanetary shocks, accounting for self-generated Alfvénic turbulence upstream of the shock. In this section, we will focus only on those features of the code, that are relevant for the present study. A comprehensive description of CSA can be found in \citet{BattarbeePHD-2013}; results and discussion of various aspects of the model are presented in a number of papers \citep{Vainio-2007, Vainio-2008, Battarbee-2011, Battarbee-2013, Vainio-2014, Afanasiev-2015}.    

CSA simulates the coupled evolution of ions and Alfv\'en waves on a single radial magnetic field line. Ions are traced under the guiding-centre approximation and Alfv\'en waves with the Wentzel–Kramers–Brillouin (WKB) approximation complemented with a term describing diffusion in parallel wavenumber, accounting for the effect of wave-wave interactions. The wave-particle interactions are treated in the framework of quasi-linear theory \citep{Jokipii-1966}, assuming the pitch-angle-independent resonance condition.  The spatial simulation domain extends from the shock front  to a large distance in the shock's upstream (usually a few hundred solar radii), i.e., the wave-particle interactions are simulated explicitly in the ambient heliospheric plasma. The effect of wave-particle interactions in the downstream is computed using a combination of test-particle simulations and calculations of the probability of return to the upstream \citep{JonesEllison-1991}. The shock itself is treated as a magnetohydrodynamic (MHD) discontinuity, the gas and magnetic compression ratios of which are computed using the Rankine-Hugoniot relations. To compute those at each time step, CSA uses analytic models of the ambient solar wind parameters (magnetic field magnitude, plasma density, solar wind speed and temperature) and of the shock parameters (shock speed and obliquity) as functions of the heliocentric distance or the simulation time. 

The model of the magnetic field implemented in CSA is given by 
\begin{equation}
B(r)=B_{0}\left(\frac{r_{\oplus}}{r}\right)^{2}\left[1+b_{f}\left(\frac{R_{\odot}}{r}\right)^{6}\right],
\label{eq:magn_field}
\end{equation}
where $r$ is distance, $R_{\odot}$ is the solar radius, $r_{\oplus}=1$~AU, and $B_{0}$ and $b_{f}$ are the parameters that have to be specified at input. The $b_{f}$ parameter allows one to account for a super-radial expansion of the associated magnetic flux tube close to the solar surface. 

The model of the plasma density is given by
\begin{equation}
n(r)=n_{2}\left(\frac{r_{\oplus}}{r+r_{1}}\right)^{2}+n_{8}\left(\frac{R_{\odot}}{r+r_{1}}\right)^{8},
\label{eq:density_model}
\end{equation}
where $n_{2}$, $n_{8}$ and $r_{1}$ are the model parameters. Here the first term accounts for the asymptotic expansion with constant solar-wind speed and the second term is the coronal component. In CSA, the solar wind speed $u_{\mathrm sw}(r)$ is computed assuming mass conservation, $n(r) u_{\mathrm sw}(r)/B(r) = \mathrm{const}$, which requires the value of the solar wind speed to be specified at some distance, e.g., at 1 AU. In this work, we take $u_\mathrm{sw}(1\mathrm{AU}) = 380~\mathrm{km\,s^{-1}}$. 

Like in some previous modelling work with CSA \citep[e.g.,][]{BattarbeePHD-2013}, we specify the cosine of the shock-normal angle $\theta_\mathrm{Bn}$ as
\begin{equation}
\mu_{\mathrm{s}}(t)=\frac{\mu_{\mathrm{s}0}+\mu_{\mathrm{s}1}\, q\, t}{1+q\, t},
\label{eq:pitch_cos}
\end{equation}
where $\mu_{\mathrm{s0}}$ and $\mu_{\mathrm{s1}}$ are its initial and asymptotic (at $t\rightarrow\infty$) values and $q$ specifies its rate of change. This form allows for modelling a shock with $\theta_\mathrm{Bn}$ decreasing with time, which is a good approximation for many CME-driven shocks. 

We also specify the shock speed along the magnetic field line as 
\begin{equation}
V_{\mathrm{s}}(t)=V_{\mathrm{s}0}(t+t_{0})^{\gamma}
\label{eq:shock_speed}
\end{equation}
where $V_{\mathrm{s0}}$, $t_{0}$ and $\gamma$ are the model parameters.

In contrast to previous studies of particle acceleration with CSA, in which some typical values were considered for the model parameters, in this work the model parameters in the above expressions were obtained by fitting corresponding data points computed from observations, using the techniques used by \citet{Rouillard-2016}. Therefore, our choice of a power-law dependence for the (field-aligned) shock speed with the additional parameter $t_0$ in Eq.~(\ref{eq:shock_speed}) as well as inclusion of the parameter $r_{1}$ in Eq. (\ref{eq:density_model}) were driven by better fitting results in the case of these models.  

Finally, we model the plasma temperature as
\begin{equation}
T(r)=T_0\left(\frac{R_{\odot}}{r}\right)^\alpha,
\label{eq:tempr}
\end{equation}
with $T_0 = 1.72\times10^6$~K and $\alpha = 0.144$, thus allowing a rather weak decrease with distance from the Sun. 
This profile is chosen ad-hoc, but in the range of radial distances of interest in this study ($1.5-6\,R_\odot$), the temperature decrease is close to the one resulting from the profile of \citet{CranBal05}. Ions and electrons are considered to have the same temperature and they both contribute to the plasma pressure. 

The seed particles are injected at the shock. In the injection procedure implemented in CSA \citep{Vainio-2007, Vainio-2008, Battarbee-2013}, the speed of a given Monte Carlo seed particle is assigned by drawing it from a distribution function, which we refer to as the seed-particle pool distribution. In the version of CSA by \citet{Battarbee-2013},  a kappa-distribution $f_\kappa(\upsilon)$ is used as the seed-particle pool distribution. It is specified by a number of parameters: the seed-particle density $n_\mathrm{seed}(r)$ and kinetic temperature $T_\mathrm{seed}(r)$, which are functions of the heliocentric distance $r$, and the constant $\kappa$ parameter, which governs the strength of the suprathermal tail. In this work, we take $T_\mathrm{seed}(r) = T(r)$, $n_\mathrm{seed}(r) = n(r)$ and $\kappa = 2$ (strong suprathermal tail). Such a low value of $\kappa$ is chosen to address one of the goals of this work: to help to understand whether or not diffusive shock acceleration in an oblique shock is able to produce protons of GLE energies. Clearly, if our model does not produce GLE protons at $\kappa = 2$, it will not produce those at higher values of $\kappa$, i.e., for lower numbers of suprathermals in the corona. Finally, in order to exclude high-energy particles from the seed population, we have additionally introduced an exponential cut-off energy $E_0$, so the seed-particle pool distribution is given by $f_\mathrm{seed}(\upsilon) = f_\kappa(\upsilon)\exp[-\tfrac{1}{2} m_\mathrm{p}\upsilon^2/E_0]$, and take $E_0 = 1$~MeV in all the simulations.   

\subsection{Deriving shock and ambient plasma parameters}
The shock and shock’s upstream plasma parameters along individual magnetic field lines are derived from the 3-D coronal shock modelling technique used in \citet{Rouillard-2016}. The shock fitting technique is based on analysis of white-light and EUV images of the corona taken by the Solar and Heliospheric Observatory (SOHO), Solar-Terrestrial Relation Observatory (STEREO) and Solar Dynamics Observatory (SDO). The magnetic field was modelled using a Potential Field Source Surface (PFSS) extrapolation\footnote
{http://www.lmsal.com/~derosa/pfsspack/}
of the photospheric magnetograms obtained by the Helioseismic and Magnetic Imaging (HMI) instrument onboard SDO. Applying the geometric fitting of the eruption images obtained from multiple vantage points, one can triangulate the 3-Dimensional (3D) expanding shock front and, correspondingly, obtain its 3D kinematics. In combination with the PFSS model, this provides the magnetic field direction and magnitude at all points on the shock surface, i.e. the 3D magnetic geometry of the shock. The white-light and EUV images were also exploited to derive the electron density in the ambient corona \citep{Zucca14}. 

An important result of \citet{Rouillard-2016} is the finding of a spatially well localised region crossing the expanding pressure/shock front and characterised by elevated (fast-mode) Mach number $M_\mathrm{FM}$ (see their Figures 4 and 5). The authors argue that this region forms just above the helmet streamer tips where the magnetic field is decreased. They associate this region with the source region of the heliospheric plasma sheet (HPS). The HPS measured in situ near 1AU has an order of magnitude lower magnetic field and higher plasma density as compared to the ambient solar wind. Moreover, their analysis shows that the release of relativistic particles occurs near the time of the most significant rise in  $M_\mathrm{FM}$.       

\subsection{Preparation of input for CSA simulations}

We obtained the magnetic field $B$ and plasma density $n$ in the ambient solar wind, upstream of the shock, as well as the shock speed along field line (determined from the PFSS model) $V_\mathrm{s}$ and the shock-normal cosine $\mu_\mathrm{s}$ for 109 individual field lines, which we then fitted using the analytic functions implemented in CSA and given by Eqs.~(\ref{eq:magn_field})-(\ref{eq:shock_speed}). 

The fitting procedure for each field line started with fitting $\mu_\mathrm{s}$ and  $V_\mathrm{s}$, both as functions of time $\tau$ after the first shock-crossing of the field line. Then,  using the fit of $V_\mathrm{s}$, the magnetic field $B$ was fitted, i.e., to fit our 1-D model given by Eq.~(\ref{eq:magn_field}) to the data, we computed the path length $s = \int_0^\tau V_\mathrm{s} d\tau$ along the field line and substituted $r$ in Eq.~(\ref{eq:magn_field}) with $s + (d_0 + d_1)$, where $d_0$ is the heliocentric distance of the shock-crossing point at $\tau=0$ and $d_1$ is an additional fitting parameter. So, there are actually three fitting parameters ($B_0$, $b_f$ and $d_1$) to fit the magnetic field data. Finally, the plasma density $n$ was fitted using Eq.~(\ref{eq:density_model}) with $r=s + (d_0 + d_1)$, but at this step $d_1$ is the fitted value obtained from the magnetic field fitting.  
  
\begin{figure*}
\centering
\includegraphics[width=.83\textwidth]{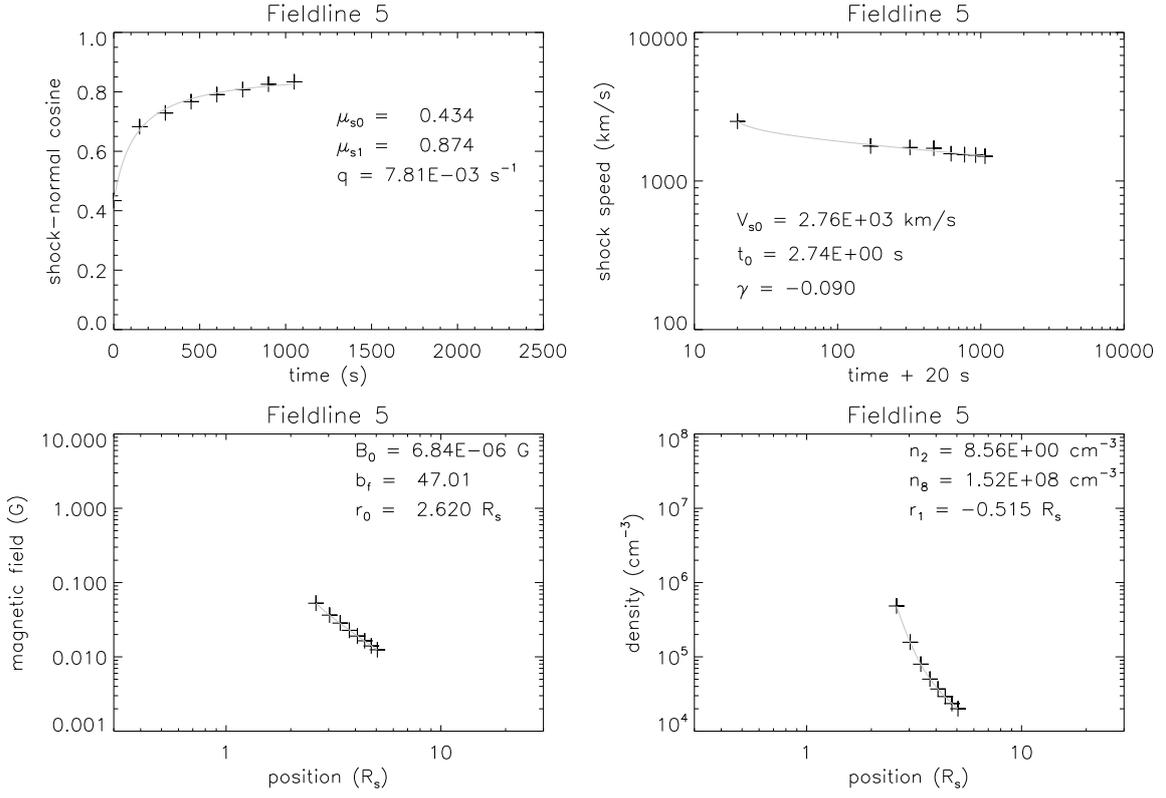}
\protect\caption{
Example of data points along a single field line, obtained by means of the semi-empirical shock modelling technique along with the corresponding fits by the analytic functions implemented in CSA. Starting from the left, the top plots show the shock-normal cosine $\mathrm{cos}\,\theta_\mathrm{Bn}$ and the shock speed along the field line $V_\mathrm{s}$, both as functions of time $\tau$ after the first shock-crossing of the field line, and the bottom plots show the magnetic field strength $B$ and the plasma density $n$, both as functions of the radial position $s + r_0$, where $r_0 = d_0 + d_1$ (see text for details). In each plot, the values of the corresponding fitting parameters are given as well.     
\label{fig:data_FL5}}
\end{figure*}

\begin{figure*}
\centering
\includegraphics[width=.83\textwidth]{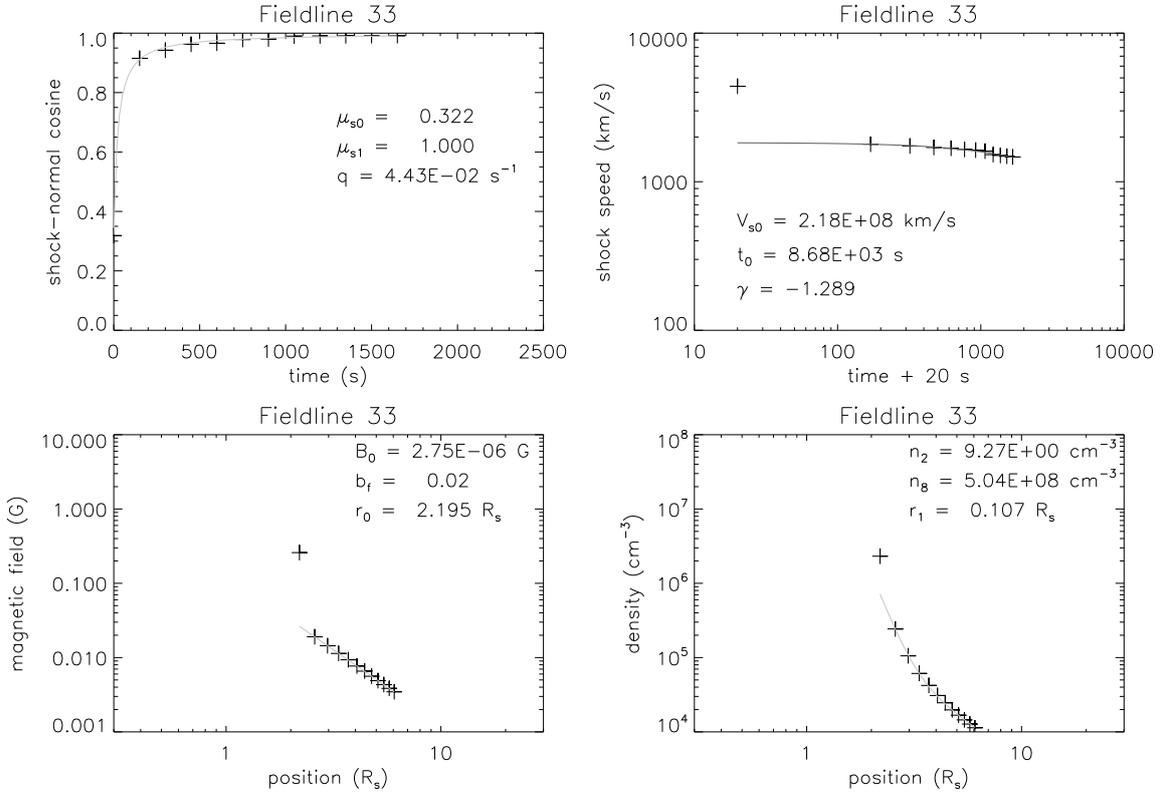}
\protect\caption{
Same as in Figure~\ref{fig:data_FL5}, but for a different field line.   
\label{fig:data_FL33}}
\end{figure*}

Next, we rejected  from further consideration those data sets, fitting of which gave unphysical values for the fitting parameters (17 data sets were rejected). Examples of the data sets with superposed fits for two different field lines are shown in Figs~\ref{fig:data_FL5} and \ref{fig:data_FL33}. Note that for many field lines the fitting results are not perfect as the first data points lie too far from the predicted positions, like in Fig.~\ref{fig:data_FL33}. Nevertheless, when choosing data sets for particle acceleration simulations, we considered such cases to be acceptable. If such a case was chosen for a simulation with CSA, we started the simulation so that the shock position at start corresponded to the second data point. In this way, we guaranteed that the plasma and shock parameters are described correctly in the simulation but, of course, this leads to some underestimation of the acceleration. 

We selected a number of field lines among the retained ones for actual simulations with CSA (see Simulation results for further details). For the selected field lines, the values of the reduced chi-square statistic computed for each of four physical quantities, assuming the standard error of $10\%$ of the measured value and excluding the first data point (for all field lines), are in the range $0.002 \leq \chi^2_\mathrm{red} \leq 0.5$.        

The real time corresponding to the starting time of the simulation ($t = 0$) for a given magnetic field is the time of intersection of that field line by the shock front or 2.5 minutes later (if the simulation has to be shifted as explained above; the time difference between data points is 2.5 minutes). Therefore, it is different for different field lines. The reference time here is the flare onset time (01:25 UT). For that reason, the corresponding initial heights of the shock front also vary from one field line to another (see Table~\ref{table:1}). 

\begin{table*} [htp]
\caption{Parameters associated with the simulated magnetic field lines.}
\centering
\begin{tabular}{r c c r r r c l l r r r r}
\hline\hline
FL & FL  & $\langle M_\mathrm{FM}\rangle$\tablefootmark{a} & $E_\mathrm{max}$\tablefootmark{b} & $E_\mathrm{c,max}$\tablefootmark{c} & $t_\mathrm{c,max}$\tablefootmark{d} & $d_{0,\mathrm{sim}}$\tablefootmark{e} & $\Delta t$\,\tablefootmark{f} & Plotted   \\        
\cline{10-13}
\#  & group &                                                & (MeV)                       & (MeV)                         & (s)                       & ($R_\odot$)  & (min) & as  \\  
\hline  
25  & W &  1.47  & 6           & 1.0         & 90         & 2.16 & 17.5$^{*}$   & black dot    \\  
72  & W &  1.96  & 6           & 4.5         & 2000     & 2.19 & 19.5             & black dash \\  
27  & W &  2.67  & 32         &  22.4 & 210    & 2.34 & 15.0$^{*}$            & black solid \\  
64  & M &  5.22  & 162        & 123.6  & 1970 & 2.60 & 17.5$^{*}$  &   green dot     \\  
1    & M &  3.22  & 315        &  237.7  & 610  & 1.57 &  12.5           &   green dash  \\  
5    & M &  4.72  &  318       &  244.4 & 490  & 2.33 &   22.5$^{*}$ &   green solid   \\  
106& S &  6.85  & 796         &  677.0   & 820    & 2.65  & 22.5$^{*}$ & red dash     \\  
60  & S &  7.24  & 848         & 723.9    & 2000  & 2.45  & 15.0$^{*}$  & red dot       \\  
33  & S &  8.66  & $>$1800 &  1433.7 & 820    & 2.55  & 15.0$^{*}$ & red solid     \\  
\hline
\end{tabular}
\tablefoot{
\tablefoottext{a}{Fast-mode Mach number averaged over 2000 seconds.}
\tablefoottext{b}{Maximum proton energy achieved in a simulation.}
\tablefoottext{c}{Maximum spectral cutoff energy achieved in a simulation.}
\tablefoottext{d}{Simulation time when the maximum cutoff energy is achieved.}
\tablefoottext{e}{Heliocentric distance of the shock front at the simulation start.}
\tablefoottext{f}{Time lag between the flare onset (01:25~UT) and the time corresponding to the simulation start (asterisk indicates that the first field line-shock intersection occurred 2.5 minutes before the time corresponding to the simulation start).}
}
\label{table:1}
\end{table*}
\section{Simulation results}\label{sec:results}

As stated above, \citet{Rouillard-2016} concluded that the region of high (fast-mode) Mach number is a favourable location for strong particle acceleration. This motivated us to investigate the effect of the Mach number in our particle acceleration simulations. The Mach number is defined as $M_\mathrm{FM} = V_\mathrm{sn}/V_\mathrm{FM,n}$, where $V_\mathrm{sn}$ is the shock-normal speed and $V_\mathrm{FM,n}$ is the fast magnetosonic wave speed along the shock normal, which is determined by the Alfv\'en speed, the sound speed and the shock-normal angle (see, e.g., \citealt{Rouillard-2016} for the exact expression).    

\begin{figure*}
\begin{centering}
\includegraphics[width=.5 \textwidth]{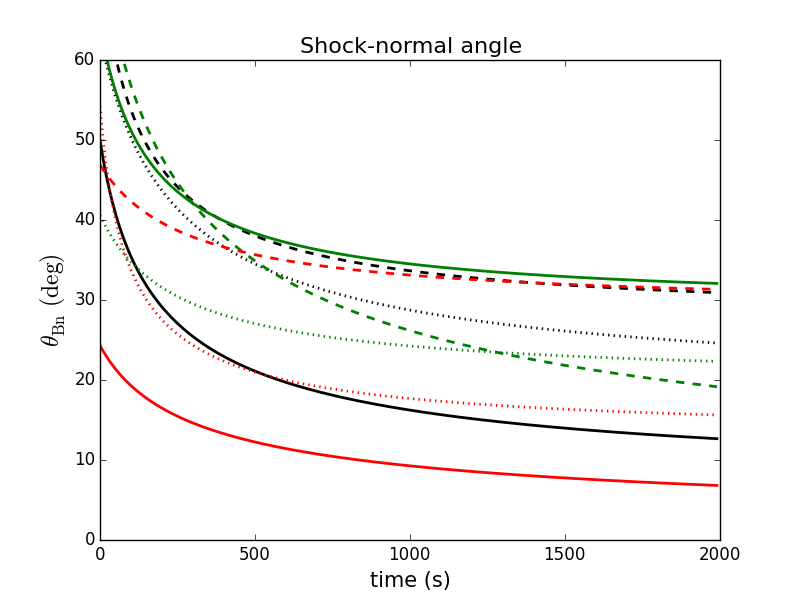}\includegraphics[width=.5 \textwidth]{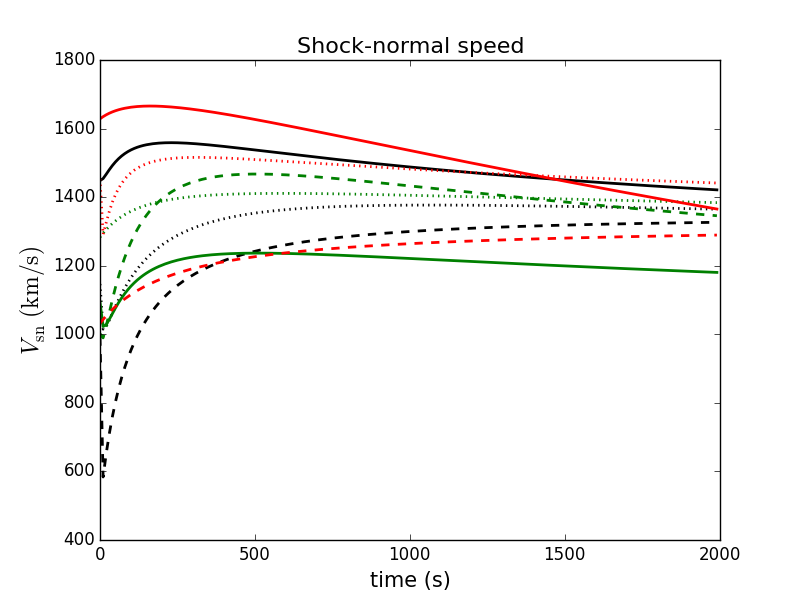}           
\includegraphics[width=.5 \textwidth]{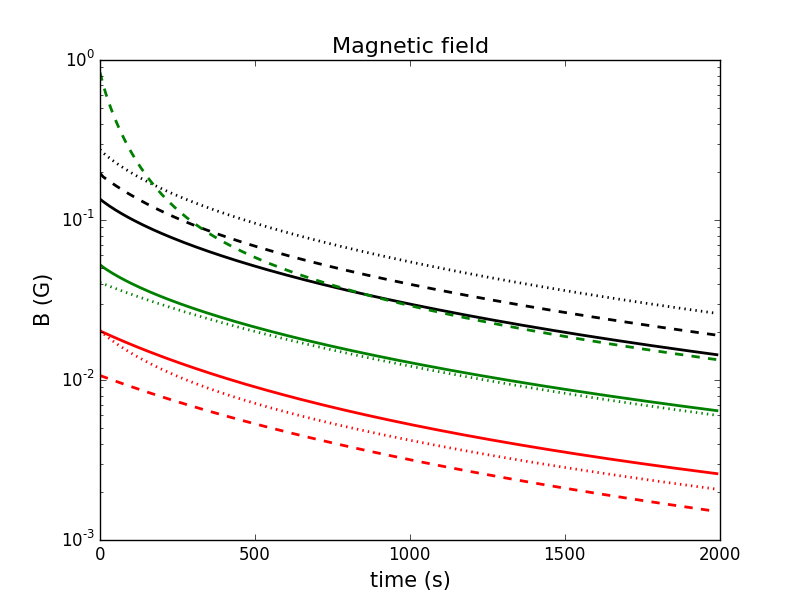}\includegraphics[width=.5 \textwidth]{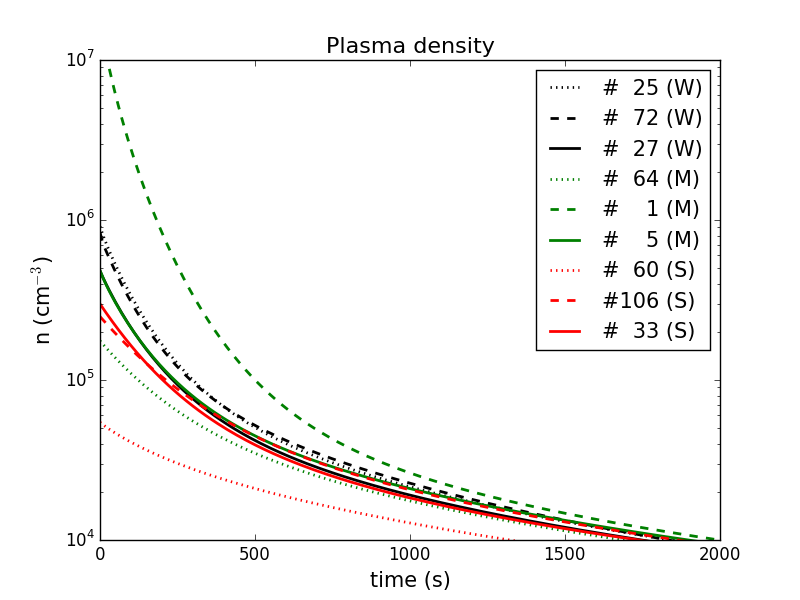}
\protect\caption{
Shock and plasma parameters resulting from fitting the data for field lines selected for CSA simulations, plotted against simulation time. The upper panels show the shock-normal angle $\theta_\mathrm{Bn}$ and the shock-normal speed $V_\mathrm{sn}$; the bottom panels show the magnetic field magnitude $B$ and the plasma density $n$. The legend gives the ID number and the group type of the simulated field lines.    
\label{fig:mach_num_compon}}
\par\end{centering}
\end{figure*}

Due to limited computing resources, we divided all the field lines pre-selected for CSA simulations into three groups according to the value of the Mach number averaged over 2000 seconds, which was the simulation time in all but one cases, and simulated three cases representing each group. The groups are the following: "weak shock" group with the average $M_\mathrm{FM} \leq 3$, "moderate shock" group with the average $3 <M_\mathrm{FM} \leq 6$, and "strong shock" group with the average $M_\mathrm{FM}>6$ (see Table~\ref{table:1}). Figure~\ref{fig:mach_num_compon} shows the plasma and shock parameters for the chosen field lines, obtained from the fitted data and plotted against simulation time (for convenience in further analysis, we show the shock-normal angle $\theta_\mathrm{Bn}$ instead of $\cos\theta_\mathrm{Bn}$ and the shock-normal speed $V_\mathrm{sn}$ instead of the field-aligned shock speed $V_\mathrm{s}$). One can see that the shock becomes less oblique in each case and $\theta_\mathrm{Bn}$ decreases faster during first $\sim300$~seconds of simulation time. Because of this (although the field-aligned shock speed $V_\mathrm{s}$ decreases as well), the shock-normal speed $V_\mathrm{sn} = V_\mathrm{s}\cos{\theta_\mathrm{Bn}}$ increases with time or has a local maximum along some field lines. The magnetic field and the density decrease monotonically with time. One can notice that (at least for the simulated field lines) the introduced group classification of the field lines correlates the best with the magnetic field strength rather than with the other parameters.

\begin{figure*}
\begin{centering}
\includegraphics[width=.5 \textwidth]{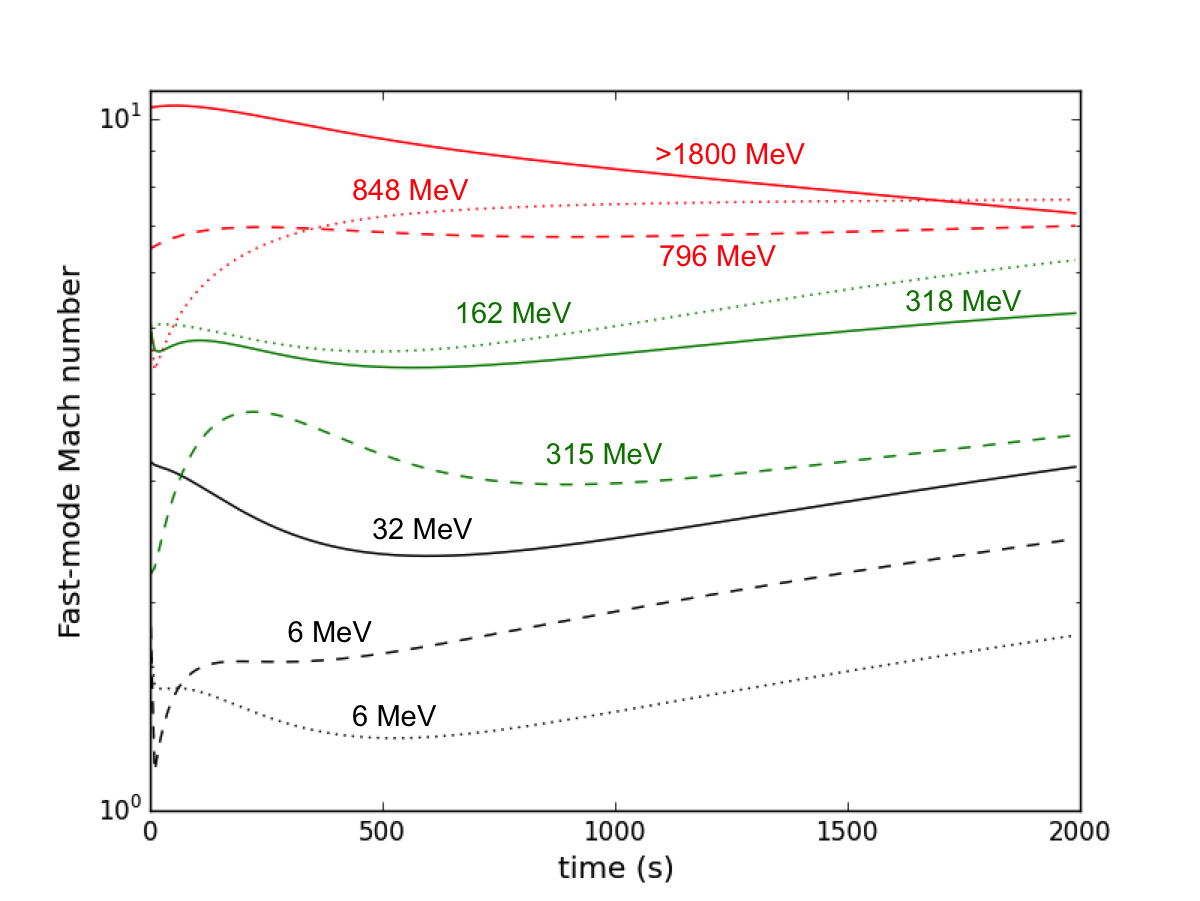}\includegraphics[width=.5\textwidth]{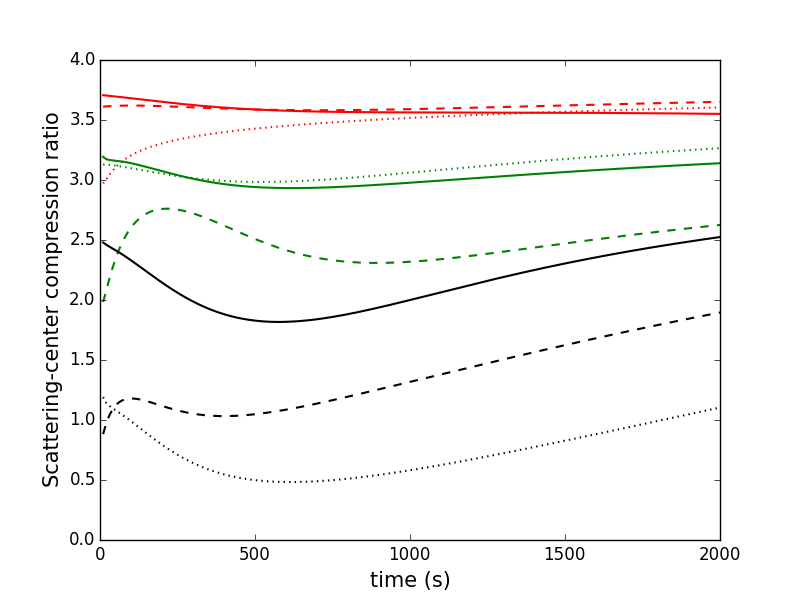}
\protect\caption{
{\it Left panel:} Fast-mode Mach number of the shock versus simulation time for 9 magnetic field lines, for which proton acceleration simulations with CSA were performed. Black curves show results for field lines belonging to the "weak shock" group of field lines, green curves the "moderate shock" group and red curves the "strong shock" group (see text for details). Numbers provide the absolute maximum proton energies attained in the simulations. The total simulation time was 2000 seconds in all the cases except the most resource-demanding case (corresponding to red solid curve), in which it was 820 seconds. {\it Right panel:} Scattering-centre compression ratio versus simulation time for the corresponding field lines.      
\label{fig:machcompr}}
\par\end{centering}
\end{figure*}

The evolution of the Mach number with simulation time along different field lines is shown on the left panel of Fig.~\ref{fig:machcompr}. The maximum proton energy $E_\mathrm{max}$ attained in each simulation is indicated for each case as well. The curves belonging to the same group of field lines are plotted with the same colour. One can see that $E_\mathrm{max}$ indeed tend to increase with the (averaged) Mach number (see also Table~\ref{table:1}). However, this tendency may not hold for the field lines from the same group (see the green group of field lines). The right panel of Fig.~\ref{fig:machcompr} shows the evolution of the scattering-centre compression ratio in the simulations. In CSA, the scattering-centre compression ratio is computed as $r_\mathrm{c}  = r_\mathrm{g}(1 - M^{-1}_\mathrm{A})$, where $r_\mathrm{g}$ is the gas compression ratio and $M_\mathrm{A}$ is the Alfv\'enic Mach number. This expression is derived under the assumption that particle scattering in the downstream region is governed by frozen-in magnetic fluctuations \citep{Vainio-2014}. Note a quite close similarity between the behaviour of $r_\mathrm{c}$ and $M_\mathrm{FM}$ in time.

\begin{figure*} 
\centering
\includegraphics[width=.5\textwidth]{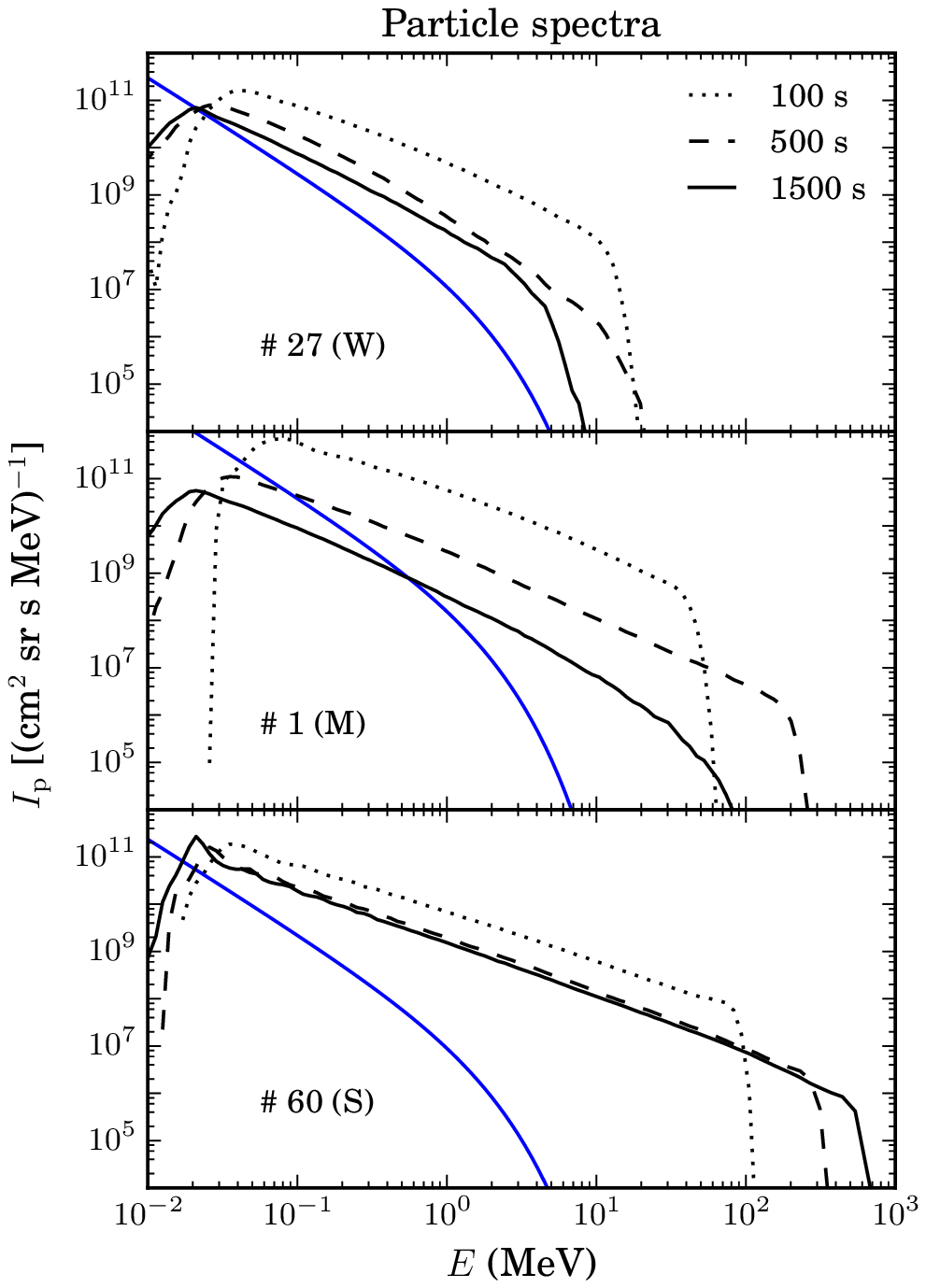}\includegraphics[width=.497\textwidth]{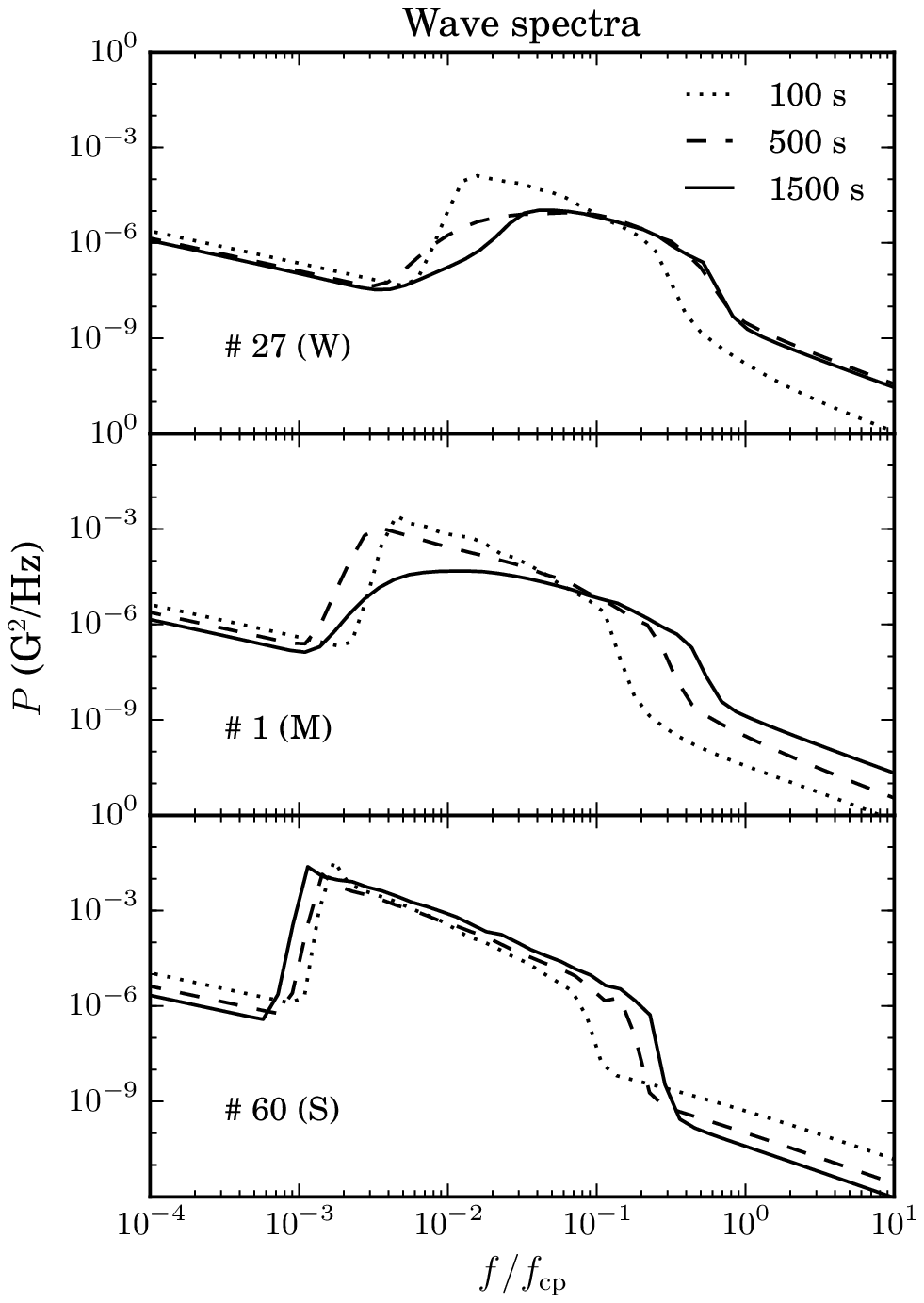}
\caption{{\it Left panel:} Simulated proton energy spectra at the shock at $t = 100$, 500 and 1500~s for 3 selected field lines. The blue line in each panel shows, for reference, the spectrum of protons constituting the seed-particle pool, $I_\mathrm{seed}(E) = p^2f_\mathrm{seed}(p)$,  at $t = 100$~s. {\it Right panel:} Corresponding simulated Alfv\'en wave power spectra $P(f)$ at the shock, plotted against  $f/f_\mathrm{cp}$, where $f_\mathrm{cp}$ is the proton cyclotron frequency.}
\label{fig:spectra_examples}
\end{figure*}

Figure~\ref{fig:spectra_examples} shows examples of simulated energy spectra of particles and power spectra of self-generated Alfv\'en waves for three selected field lines, each representing a particular field line group, corresponding to three moments in time. The wave spectra are plotted against the ratio $f/f_\mathrm{cp}$, where $f_\mathrm{cp}$ is the proton cyclotron frequency. In this way, we exclude the systematic shift of the bulk of the amplified spectrum to lower frequencies, resulting from the systematic decrease of the large-scale magnetic field since $f_\mathrm{res} = f_\mathrm{cp} V/\upsilon$, and can see better the effect of particle acceleration. 

As expected, the simulated particle spectra have a clearly visible power-law part and a rollover. On the other hand, the spectra corresponding to different field lines evolve differently in terms of both the power-law index and the rollover. 
Moreover, it can be noticed that the number of high-energy particles in the spectrum for FL~27 decreases between $t = 500$ and 1500~s (this is also true for FL~1, but cannot be seen from the plot). This is accompanied by a decrease in the wave energy at the resonant frequencies.    

\begin{figure*}
\centering
\includegraphics[width= .5\textwidth]{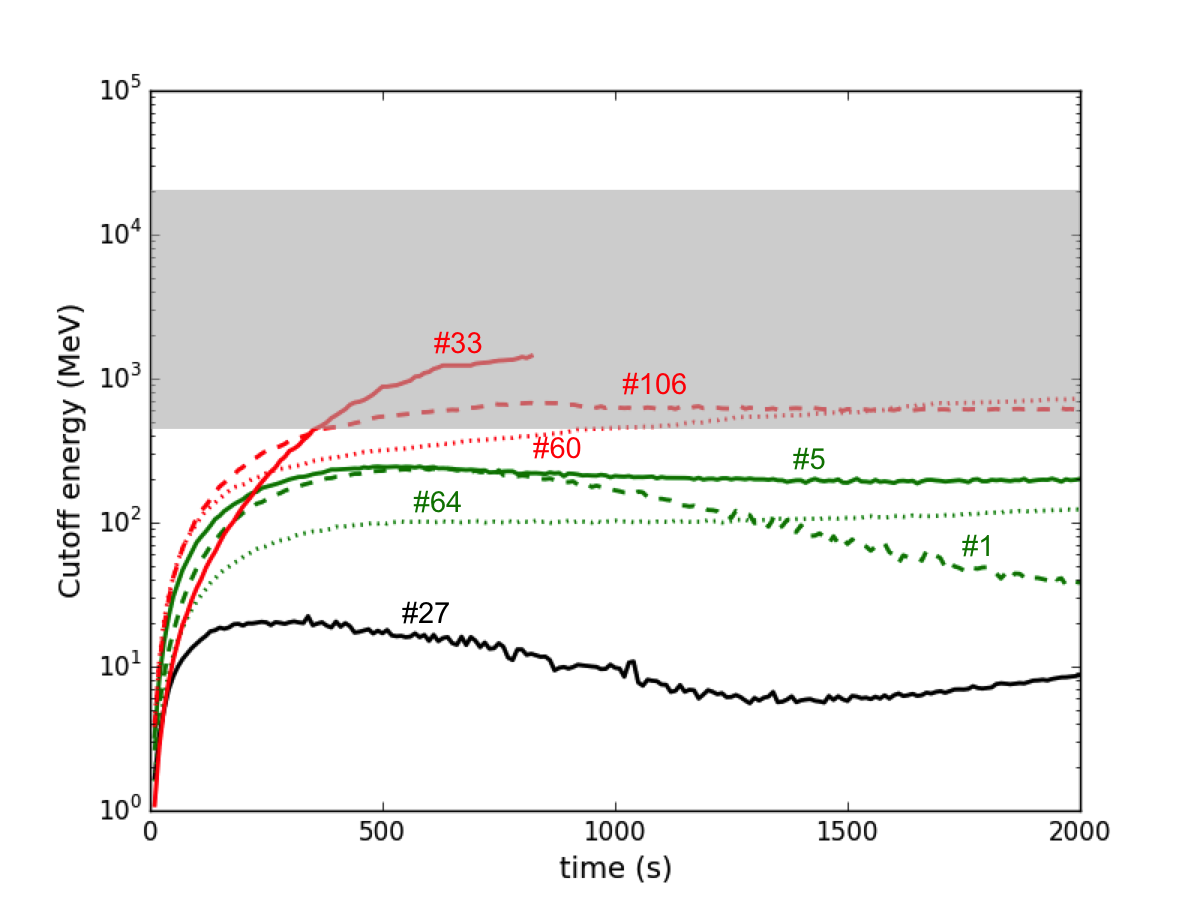}\includegraphics[width= .5\textwidth]{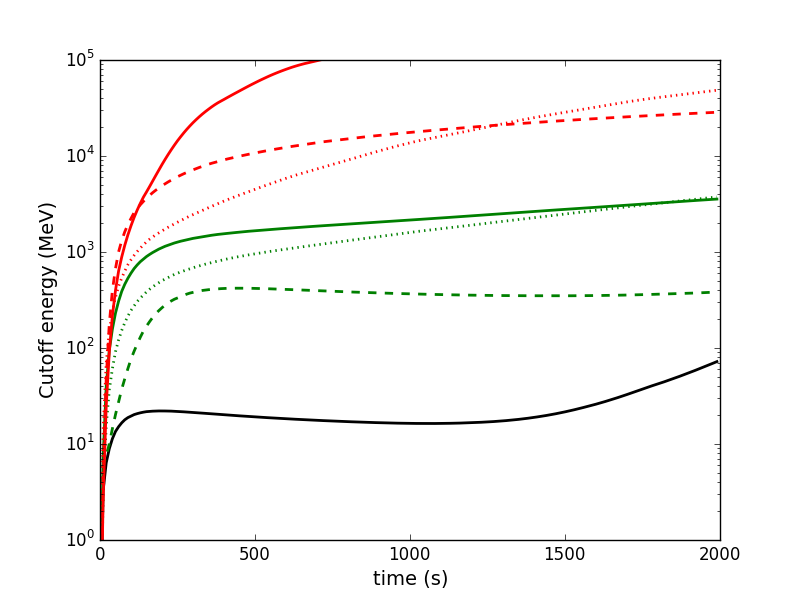}
\caption{
{\it Left panel:} Simulated spectral cutoff energy versus time for different field lines. The shaded area shows the proton energy range corresponding to GLE energies. {\it Right panel:} Numerical solutions of Eq.~(\ref{paral_acc_rate}) for the selected field lines.    
\label{fig:max_energy_sum}}
\end{figure*}

To study how acceleration efficiency of the shock evolves with time in the simulations, we use the cutoff energy $E_\mathrm{c}$ of the particle spectrum at the shock. It is often determined by fitting the spectrum by a function $CE^{-\gamma}\exp[-(E/E_\mathrm{c})^\delta]$, where $C$, $\gamma$, $\delta$ and $E_\mathrm{c}$ are the fitting parameters \citep[see, e.g.,][]{Afanasiev-2015}. There are also other methods that allow one to determine the cutoff energy faster but less accurately. For instance, \citet{Vainio-2007} defined the cutoff energy as the point of intersection between the simulated spectrum and the theoretical spectrum of \citet{Bell-1978} divided by 10. Although this method provides only an approximation for the magnitude of the actual cutoff energy (given by the fitting method), it still allows tracking dynamical changes of the spectrum in the course of a simulation. In this work, we define the cutoff energy as the point of intersection between the simulated particle spectrum at the shock and the power-law function first fitted to the simulated spectrum in the energy interval from 1 to 10 MeV and then divided by 10.  
In the cases where the power-law part of the spectrum clearly did not extend beyond 10 MeV (e.g., in the beginning of a simulation), we used as a cutoff energy the point of intersection between the simulated spectrum and the seed-pool particle spectrum. The {\it left panel} of Fig.~\ref{fig:max_energy_sum} shows the evolution of the cutoff energy in the simulations with time for different field lines. For four field lines (FLs~1, 5, 27, and 106), the cutoff energy reaches the maximum within $\sim800$~seconds of the simulation, followed by a decrease (the most pronounced for FLs~1 and 27) and a secondary increase (FL~27). For FLs~60 and 64, the cutoff energy monotonically increases during almost the whole simulation time of 2000~seconds. For FL~33, there is a monotonic increase in the cutoff energy until the end of the simulation at $t = 820$~s. We do not show here the results for FL~25 and FL~72 as the inferred cutoff energy is $\leq 1$~MeV (for FL~72, during the first half of the simulation). In fact, there are time intervals during which the scattering centre compression ratio for these field lines is $\leq 1$ (see Fig.~\ref{fig:machcompr}, right panel) and, hence, DSA does not operate. The injection of particles at those times is provided by the shock drift acceleration mechanism. 

\section{Discussion}\label{sec:discussion}

We shall start with discussion of the evolution of particle energy spectra in our simulations (Fig.~\ref{fig:spectra_examples}). In the steady-state DSA theory of Bell, the power-law index of the spectrum $\propto E^{-\gamma}$ is determined by the scattering-centre compression ratio $r_\mathrm{c}$ of the shock as $\gamma = (r_\mathrm{c}+2)/[2(r_\mathrm{c}-1)]$ \citep[for non relativistic energies, e.g.,][]{Vainio-2014}. Increase of $\gamma$ in a weakening shock explains naturally the development of a more gradual spectral rollover (like in the spectrum for FL~27 at $t=500$~s in Fig.~\ref{fig:spectra_examples}). In a non-steady-state situation, if the shock parameters evolve slowly, the spectral index at a given time can be expected to be approximately determined by the instantaneous value of $r_\mathrm{c}$. For each simulated field line, we fitted a power law to the simulated spectrum at each time and then compared the resulting spectral index with $\gamma$. We have seen that the evolution of the power-law spectral index closely follows the evolution of $r_\mathrm{c}$. This can also be seen from comparison of the particle energy spectra in Fig.~\ref{fig:spectra_examples} with the evolution of the scattering-centre compression ratio on the right panel of Fig.~\ref{fig:machcompr}.  

We also would like to understand the temporal behaviour of the particle cutoff energy along different field lines. The decreasing cutoff energy implies that there is some energy/particle loss mechanism at work. A detailed examination of the time series of particle energy spectra shows that the implied mechanism affects mainly high-energy particles (this can also be seen in Fig.~\ref{fig:spectra_examples} for FL~27). This indicates that the wave-particle interactions alone cannot account for the cutoff energy decrease because a change $d\mu$ in the particle's pitch-angle cosine $\mu$ due to pitch-angle scattering (as measured in the wave frame) results in a relative momentum change (as measured in the solar-fixed frame) $dp/p = (V/v)\,d\mu$, where $V = u_\mathrm{sw} + V_\mathrm{A}$ is the phase speed of Alfv\'{e}n waves and $v$ is the particle speed, both with respect to the solar-fixed frame. Therefore, the relative momentum change due to pitch-angle scattering is larger for lower-energy particles.

\begin{figure}[tbh]
\centering
\includegraphics[width= .5\textwidth]{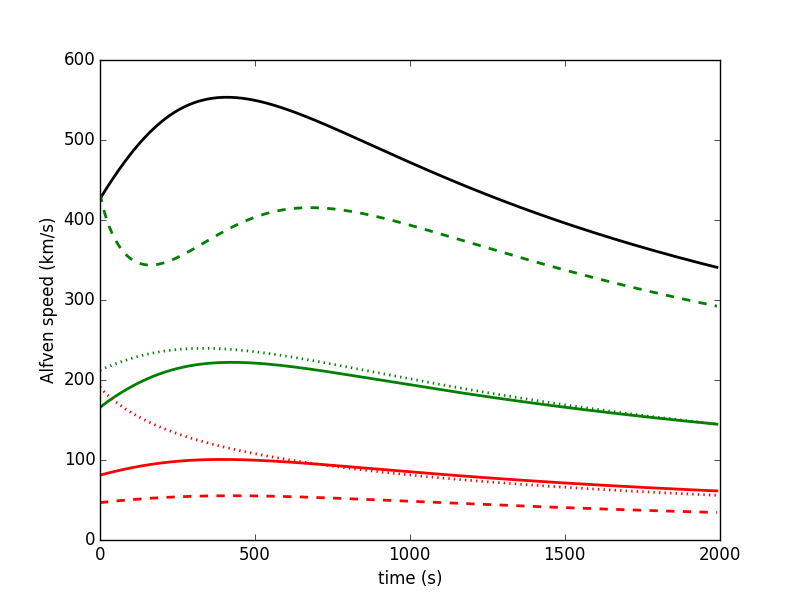}
\caption{Alfv\'en speed versus simulation time for the field lines addressed in Fig.~\ref{fig:max_energy_sum}.
\label{fig:alfven_speed}}
\end{figure}

The mechanism of interest can be associated with the weakening of the turbulent foreshock due to a local maximum in the Alfv\'en speed-radial distance profile in the corona. It was shown by \citet{Vainio03} that when the shock propagates towards a decreasing $V_\mathrm{A}$, particle trapping in the foreshock should become less efficient and the flux of escaping particles should increase. In this case, as the escaping flux has a harder spectrum than that of the trapped particles, high-energy particles can escape more efficiently, which could lead to erosion of the high-energy part of the source spectrum. Figure~\ref{fig:alfven_speed} shows the Alfv\'en speed plotted against simulation time for different field lines. One can see that most of the field lines have a maximum in the Alfv\'en speed in the time interval from 300 to 800 seconds. In addition, adiabatic cooling of particles, caused by the flux tube expansion and affecting particles of all energies, can also play a role.  

Comparison of the Alfv\'en speed profiles with the simulated cutoff energy profiles ( in Fig.~\ref{fig:max_energy_sum}) for FL~1 (dashed green line) and FL~27 (solid black line) shows that in the case of FL~1 the maximum cutoff energy is a bit delayed ($t = 610$~s) with respect to the maximum of the Alfv\'en speed ($t = 580$~s), whereas in the case of FL~27 it occurs somewhat earlier (at $t \approx 210$~s) than the maximum of the Alfv\'en speed ($t = 410$~s). Therefore, we could expect that in the case of FL~1 the decrease in the cutoff energy is mainly caused by the weakening of the turbulent trap, and in the case of FL~27 both mechanisms are at work. FL~106 (dashed red line) experiences some reduction of the cutoff energy, and also FL~5 (solid green line) but much smaller.

In order to better understand how the acceleration process proceeds in our simulations and which plasma and shock parameters control it, we combine the steady-state DSA theory of Bell with the temporal dependence of the cut-off energy in the particle energy spectrum. This approach was used earlier by \citet{Vainio-2014} to construct a time-dependent semi-empirical model of the foreshock. We write the proton acceleration rate in the following form:  
\begin{equation}
 \frac{1}{p} \frac{dp}{dt} = \frac{\pi}{2} \epsilon \Omega_\mathrm{p} \left(M_\mathrm{A} - 1\right)\,\left(\frac{p}{p_\mathrm{inj}}\right)^{-\frac{3}{r_\mathrm{c} - 1}} - \frac{1}{3}\left( \frac{V}{L} + \frac{dV}{dr}\right),
\label{paral_acc_rate}
\end{equation} 
where the first term on the right-hand side results from Eq. (15) of \citet{Vainio-2014} and the second term introduces the effect of adiabatic cooling into consideration. Here $\Omega_\mathrm{p}=eB/m_\mathrm{p}$ is the proton cyclotron frequency and it is assumed that protons being injected into the acceleration process constitute a fraction $\epsilon$ of upstream protons and the injection occurs at  $p = p_\mathrm{inj}$; $L = -B/(dB/dr)$ is the focusing length. One can see that the main controlling factor in the first term is the power-law factor, which is determined by the scattering-centre compression ratio.

Using Bell's solution for the particle distribution function at the shock,
\begin{equation}
f_\mathrm{s}(p) = \frac{\sigma\epsilon n}{4\pi p_\mathrm{inj}^3} \left(\frac{p}{p_\mathrm{inj}}\right)^{-\sigma},
\end{equation}   
where $\sigma = 3r_\mathrm{c}/(r_\mathrm{c}-1)$, the injection efficiency $\epsilon$ can be given as
\begin{equation}
\epsilon = \frac{4\pi p_\mathrm{inj}}{\sigma n} j_\mathrm{s}(p_\mathrm{inj}),
\label{bell_epsilon}
\end{equation}     
where $j_\mathrm{s}(p_\mathrm{inj})$ is the particle intensity at the shock at $p = p_\mathrm{inj}$. 

We integrated Eq.~(\ref{paral_acc_rate}) numerically for each field line to see how well this equation describes the acceleration processes along different field lines. In the integration, for the parameters $u_1$, $n$ and $B$, we used values extracted from the simulations.  
The injection efficiency $\epsilon$ was computed using Eq.~(\ref{bell_epsilon}). The parameters $p_\mathrm{inj}$ and $j_\mathrm{s}(p_\mathrm{inj})$ entering this equation are governed in the simulations by the population of particles that get back to the upstream region after the first encounter with the shock. Therefore, these parameters were taken to be the momentum and intensity corresponding to the maximum of particle energy spectrum at the shock (these parameters evolve with time as well). The integration was done with the initial condition $p|_{t=0} = p_\mathrm{inj}|_{t=0}$.
\begin{figure*}[tbh]
\centering
\includegraphics[width=.5\textwidth]{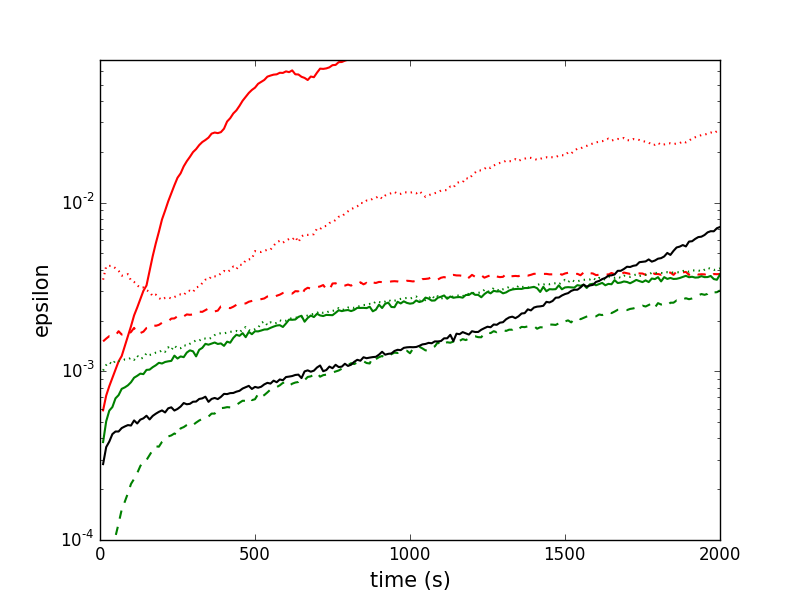}\includegraphics[width=.5\textwidth]{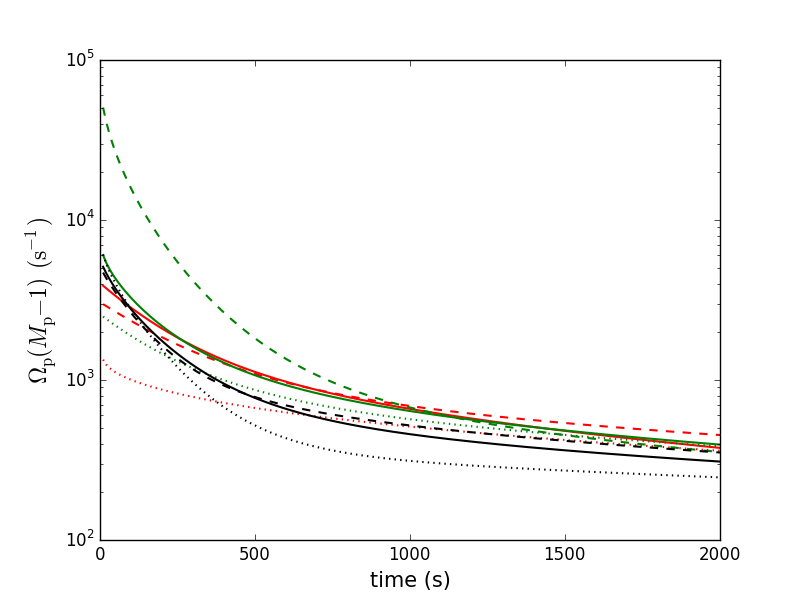}
\caption{
Injection efficiency $\epsilon$ obtained with Eq.~(\ref{bell_epsilon}) ({\it left panel}) and the parameter $\Omega_\mathrm{p}(M_\mathrm{A} -1)$ ({\it right panel}) as a function of simulation time for different field lines.    
\label{fig:eq_params}}
\end{figure*}

The solutions of Eq.~(\ref{paral_acc_rate}) (given for energy instead of momentum) are shown on the right panel of Fig.~\ref{fig:max_energy_sum}. By comparison with Fig.~\ref{fig:machcompr}, one can conclude that the numerical solutions are governed mainly by the scattering-centre compression ratio $r_\mathrm{c}$. Furthermore, the numerical solutions reproduce the time series of $E_\mathrm{c}$ obtained in the simulations quite well. This indicates that $r_\mathrm{c}$ is still the main parameter governing the particle acceleration in our simulations. However, there are also discrepancies, e.g., with respect to the initial increase rate and the attained maximum level (with the exception of FL~1 (dashed green line) and FL~27 (solid black line)). These discrepancies stem from the assumption underlying Eq.~(\ref{paral_acc_rate}) that the foreshock is in a fully developed steady state at any time, which implies better particle acceleration conditions than those realising in the actual simulations (see \citet{Vainio-2014} for discussion on this matter). One can expect that such discrepancies should be larger for stronger shocks, which is indeed the case (cf. the scattering-centre compression ratios in Fig.~\ref{fig:machcompr}). Furthermore, the decrease in the cutoff energy is smaller in the solutions obtained by integration. It can be seen rather well only for FL~27. This favours our suggestions that the decrease in the case of FL~1 is due to the turbulent trap weakening and in the case of FL~27 due to both adiabatic deceleration and turbulent trap weakening.       

We shall neglect the adiabatic cooling term in Eq.~(\ref{paral_acc_rate}) due to its smallness in the following analysis, and examine the injection efficiency $\epsilon$  and a factor $\Omega_\mathrm{p}(M_\mathrm{A} -1)$ entering the equation for each simulated field line (see Fig.~\ref{fig:eq_params}). 

According to \citet{Battarbee-2013}, who carried out a detailed study of the particle injection model implemented in CSA, the injection efficiency increases with decreasing shock-normal angle $\theta_\mathrm{Bn}$, especially strongly at small $\theta_\mathrm{Bn}$ (see, e.g., their Fig.~6). We see this feature in our simulations. Specifically, FL~33 (solid red line), which takes the lowest values of $\theta_\mathrm{Bn}$, reaches the largest values of $\epsilon$, whereas FL~1 (dashed green line), which is the most oblique during first 350 seconds in the simulation, has the lowest $\epsilon$ (cf. Figs.~\ref{fig:eq_params} and \ref{fig:mach_num_compon}). Besides, one can see that there is a dependence on the shock strength given by $r_\mathrm{c}$ or $M_\mathrm{FM}$  (cf. Fig.~\ref{fig:machcompr}).  

Furthermore, results of \citet{Battarbee-2013} indicate that the range of $\theta_\mathrm{Bn}$, in which the injection efficiency strongly increases, should extend to larger values of $\theta_\mathrm{Bn}$ for larger values of the shock-normal speed. Namely, in their simulation for $V_\mathrm{sn} = 1500~\mathrm{km\,s^{-1}}$, the injection efficiency increases by two orders of magnitude for $\theta_\mathrm{Bn}$ decreasing from $7.5^\circ$ to $0^\circ$, and for $V_\mathrm{sn} = 2000~\mathrm{km\,s^{-1}}$ a similar increase is obtained for $\theta_\mathrm{Bn}$ decreasing from $15^\circ$ to $0^\circ$. In our simulations, $\theta_\mathrm{Bn}$ changes from $24^\circ$ to $10^\circ$ within the whole simulation for FL 33 (within 820 seconds of simulation time). These values are larger than those obtained by \citet{Battarbee-2013}. Note, however, that in our simulations the shock corresponding to this field line is much stronger than in their study. Also, it follows from their Fig.~6 that if the shock is quite oblique ($\theta_\mathrm{Bn} > 20^\circ$) than the injection efficiency clearly decreases with increase in the shock-normal speed. Therefore, we can conclude that the shock's strength, obliquity and speed are all significant in controlling the injection.  

On the right panel of Fig.~\ref{fig:eq_params} we present the parameter $\Omega_\mathrm{p}(M_\mathrm{A} -1) \propto u_1\sqrt{n} (1 - M_\mathrm{A}^{-1})$. One can see that FL~1 (dashed green line) is characterised by the largest values of this parameter during the first 500 seconds of simulation time, which is due to enhanced plasma density and high obliquity of the shock along this field line. Hence, this parameter makes a larger impact on the numerical solution of Eq.~(\ref{paral_acc_rate}) for this field line (Fig.~\ref{fig:max_energy_sum}). Moreover, this finding explains why in the actual simulation for this field line the maximum/cutoff energy is almost as high as for FL~5 (solid green line), but is not reduced that much if compared with the corresponding numerical solution. A possible reason is that the significantly enhanced upstream plasma density provides an enhanced flux of injected particles that quickly amplifies the Alfv\'en wave spectrum to high intensities, comparable to the theoretical ones. Note that the particle intensity at injection energy is indeed larger for this field line than for others in the interval from \textasciitilde30 to 200 seconds (compare also the maxima of the particle spectra along different field lines in Fig.~5 at $t = 100$~s). This explanation is also supported by the fact that the acceleration rate in the beginning of the simulation is not reduced much compared to the numerical solution. Thus, the upstream plasma density can strengthen the acceleration efficiency of the shock.

\begin{figure}[tbh]
\centering
\includegraphics[width=.5\textwidth]{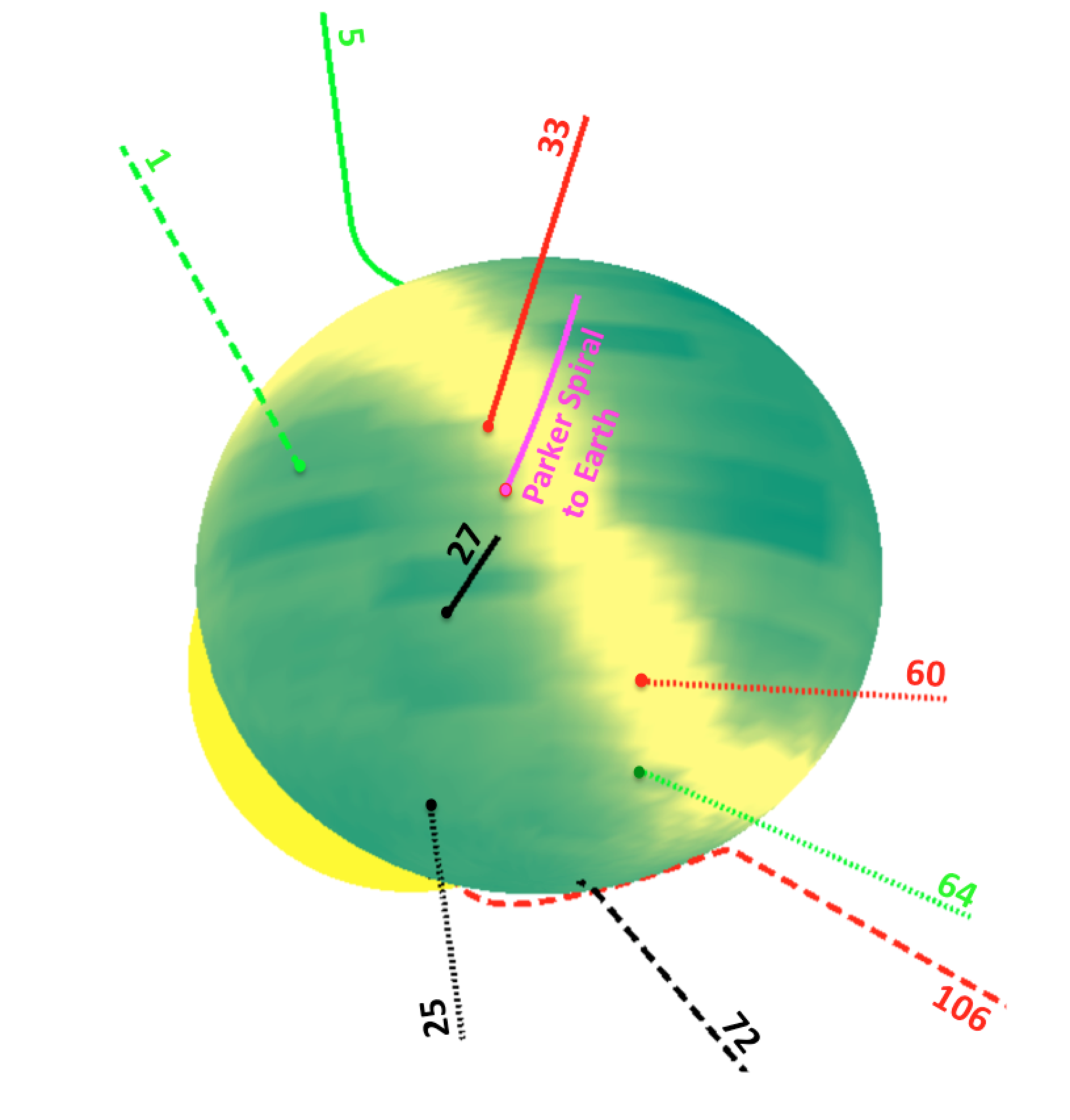}
\caption{Distribution of the Mach number over the shock front at 01:45~UT (20 minutes after the flare onset) crossing the simulated magnetic field lines. Note the band (light yellow) of high values of the Mach number.
\label{fig:3Dshock}}
\end{figure}

Figure~\ref{fig:3Dshock} shows the Mach number distribution over the shock front at 01:45~UT together with the simulated field lines. It can be seen that the field lines belonging to the "strong shock" group (red lines) indeed cross the band of high Mach number. By the indicated time, according to our simulations, protons accelerated along FL~33 (solid red line) have reached the rigidity of 1~GV. This is the earliest production of relativistic protons after the flare onset, \textasciitilde20 minutes, in comparison with the other two field lines (27.5 minutes for FL~106 and 31.5 minutes for FL~60). Of these 20 minutes, the acceleration process itself to 1~GV takes only \textasciitilde5 minutes (Fig.~6) and the other 15 minutes is the lag between the flare onset and the time corresponding to the simulation start (Table~1). The inferred relativistic proton production time (01:45~UT) is \textasciitilde5 minutes later than the SPR time (01:40~UT) determined by \citet{Gopal13} and \textasciitilde8.6 minutes later than the SPR time (01:37:20~UT) inferred by \citet{Rouillard-2016}. Note, however, that this field line is rather far from the CME origin on the Sun as it takes \textasciitilde12.5 minutes (after the flare onset) for the shock to reach it. There are several field lines in our total set of 109 lines, which are located closer to the CME origin, so that their first shock intersection occurs earlier (at 01:32:30~UT, the earliest available time). Those field lines are characterised by high values of the Mach number, comparable with those obtained for FL~33, starting from 01:37:30~UT (i.e., 5 minutes after the first intersection, recall that the data are available at 2.5 minute time resolution). Those field lines were not taken for the acceleration simulations as the quality of the parameter fitting was not good enough. Nevertheless, assuming the acceleration conditions along those field lines as favourable as for FL~33 staring from 01:37:30~UT, we should get 1~GV protons 5 minutes later, i.e., at 01:42:30~UT. This is still 2.5 minutes later than the SPR time of \citet{Gopal13} (at the error bar margin). 

On the other hand, for many field lines (including FL~33) we removed the initial 2.5-minute interval following the time of first intersection from the simulations (see Table~1), during which the shock is, however, weaker than at later times. So, perhaps, we somewhat underestimate the acceleration rate for those field lines. Also, the above analysis indicates that we need the shock and plasma parameters with better time resolution.  

Therefore, it could be that the relativistic proton production in the CME-driven shock occurs at Gopalswamy et al. SPR time (01:40~UT). Let us recall, however, that this SPR time is obtained by using neutron monitor data and {\it assuming} the path length travelled by the particles from the source on the Sun to the Earth. But, it is rather unlikely, unless there are even more favourable conditions, that our proton acceleration model can produce 1~GV protons in this event at Rouillard et al. SPR time (01:37:20~UT), which was inferred from the velocity dispersion analysis with the use of SEP data. This may indicate that quasi-perpendicular portions of the shock play the major role in particle acceleration to relativistic energies at least in this GLE event \citep{SandroosVainio09}. This also indicates importance of accurate determination of the SPR time.     

In that connection, it is worth noting that the delays obtained for our relativistic proton productive field lines are in good correspondence with the delays obtained for 16 GLE events in cycle~23 with the average delay being 24.4 minutes \citep{Gopal12}. In terms of the heliocentric distances, the production of relativistic protons in our simulations occurs at $r_\mathrm{rel} = 3.3\,R_\odot$ for FL~33 and FL~106 and at $r_\mathrm{rel} = 4.6\,R_\odot$ for FL~60. These numbers are among those obtained for GLE events in cycle~23 \citep{Reames09}.    
     
\section{Conclusions}\label{sec:conclusions}
In this work, we have simulated proton diffusive shock acceleration, using results of semi-empirical modelling of the CME-driven shock associated with the 17 May 2012 GLE event. In contrast to previous studies combining semi-empirical/MHD shock models with test-particle DSA models, we employ the Coronal Shock Acceleration simulation model, which simulates interactions of protons with Alfv\'en waves in the upstream region of the shock. This model allows simulations of the proton acceleration along individual magnetic filed lines. 

Based on the analysis of the simulations for nine magnetic field lines, we have found that the acceleration efficiency of the shock, i.e., its ability to accelerate particles to high energies, tends to be higher for those shock portions that are characterised by larger values of the scattering-centre compression ratio $r_\mathrm{c}$/fast-mode Mach number $M_\mathrm{FM}$. By comparison with Bell's steady-state DSA theory supplemented with the temporal dependence of the spectral cutoff energy, we concluded that $r_\mathrm{c}$ is the main shock parameter controlling the acceleration process in our simulations, but $M_\mathrm{FM}$ is sensitive to the magnitude of $r_\mathrm{c}$ and reveals quite closely its evolution in time. Our results, therefore, support the expectation of \citet{Rouillard-2016} that most of the acceleration occurs in the region of increased Mach number. At the same time, the acceleration efficiency can be strengthened due to enhanced plasma density in the corresponding flux tube. Our acceleration model also provides enhanced acceleration efficiency at shocks having small shock-normal angle $\theta_\mathrm{Bn}$.  

We have also found that the cutoff energy in the particle energy spectrum at the shock starts to decrease with time along some field lines. We considered two possible mechanisms leading to this effect: adiabatic cooling, which provides loss of energy for particles, and weakening of the turbulent trap due to a local maximum in the Alfv\'en velocity - radial distance profile, which provides loss of trapped particles. Our calculations show that the energy loss provided by the effect of adiabatic cooling alone is too small to explain the decrease of the cutoff energy. This suggests that both mechanisms may contribute with a larger contribution from the trap weakening mechanism.   

Our simulations show the production of protons of GLE energies for some field lines, even though the particle acceleration efficiency of the shock in the simulations is significantly lower along those field lines than if calculated based on the DSA theory.

We have inferred the delays between the flare onset and the 1 GV proton production times for different field lines in our simulations, and compared those with the delays between the flare onset and the SPR times derived from the observations for this GLE event by \citet{Gopal13} and by \citet{Rouillard-2016}. If the SPR time obtained by \citet{Rouillard-2016} is taken as the true one, it is rather unlikely that our model is able to explain the relativistic proton production in this event. This may indicate that a possibility that quasi-perpendicular portions of the shock play the main role in producing relativistic protons.  

In this paper, we focused the acceleration of SEPs during the first 20 minutes of a CME's 3-D expansion. In future studies, we will investigate the onset of SEP events measured over heliocentric longitudinal bands that exceed 180 degrees \citep[e.g.,][]{Rouillard12, Lario17}. These SEPs have been related to the lateral expansion of CMEs and their shocks propagating over extended regions of the corona. However the shock properties that control particle energisation remain unclear and our numerical model could shed new light on these events.   

\begin{acknowledgements}
This project has received funding from the European Union’s Horizon 2020 research and innovation programme under grant agreement No. 637324 (HESPERIA) and from the Academy of Finland (project 267186). The computer resources of the Finnish IT Center for Science (CSC) and the FGCI project (Finland) are acknowledged. A. Aran has been partially supported by the Spanish Ministerio de Economía y Competitividad, under the project AYA2013-42614. M. Battarbee wishes to thank the Leverhulme Trust for providing funding through grant number RPG-2015-094. A.P. Rouillard acknowledges funding from the Agence Nationale de la Recherche for the COROSHOCK project (ANR-17-CE31-0006-01).     
\end{acknowledgements}

\bibliographystyle{aa}

\end{document}